\documentclass[reprint,prl,showpacs,amsmath,amssymb,floatfix]{revtex4-1}
\usepackage{graphicx}
\usepackage{amssymb}
\usepackage{epstopdf}
\usepackage{bm}
\usepackage{pstricks}

\renewcommand{\exp}{e^}
\renewcommand\Re{\operatorname{Re}}
\renewcommand\Im{\operatorname{Im}}
\newcommand\arcsinh{\operatorname{arcsinh}}

\begin{document}
\title{Long-Range Superharmonic Josephson Current}
\author{Luka Trifunovic}
\affiliation{Department of Physics, University of Basel, Klingelbergstrasse 82,
CH-4056 Basel, Switzerland}
\date{\today}

\begin{abstract}
We consider a long superconductor-ferromagnet-superconductor junction with one
spin-active region. It is shown that an \textit{odd} number of Cooper pairs
cannot have a long-range propagation when there is \textit{only one} spin-active
region. When temperature is much lower than the Thouless energy, the coherent
transport of \textit{two} Cooper pairs becomes dominant process and the
\textit{superharmonic} current-phase relation is obtained
($I\propto\sin2\phi$).
\end{abstract}

\maketitle

The interplay between superconducting and ferromagnetic ordering has been
the subject of intensive theoretical and experimental
research~\cite{buzdin_proximity_2005,chtchelkatchev_0_2001,
ryazanov_coupling_2001,*kontos_inhomogeneous_2001,bergeret_odd_2005,
houzet_long_2007,braude_fully_2007,*asano_josephson_2007,*eschrig_triplet_2008,
khaire_observation_2010, volkov_odd_2010,*trifunovic_long-range_2010,
sellier_half-integer_2004}. It has been predicted that the hybrid systems
containing superconductors (S) and ferromagnets (F) allow the realization of
the Josephson $\pi$-junctions,
$I\propto\sin(\phi+\pi)$~\cite{buzdin_proximity_2005}. The spectral
decomposition of Josephson current-phase relation (CPR) gives
$I=I_1\sin\phi+I_2\sin2\phi\dots$. At the transition between the $0$ and the
$\pi$ phases, $I_1$ vanishes, and it is possible to obtain dominant second
harmonic ($I_2$) in the CPR. Unfortunately, the $0$-$\pi$ coexistence is very
sensitive to the temperature changes and interface
roughness~\cite{chtchelkatchev_0_2001}.

The experimental realizations of $\pi$-junctions remained elusive for a long
time; the breakthrough came with the fabrication of \textit{weak}
ferromagnets~\cite{ryazanov_coupling_2001,*kontos_inhomogeneous_2001}. Indeed,
it has been recognized that the proximity effect in a ferromagnet is
short-ranged. The electron and hole excitations acquire a nonzero relative phase
in the ferromagnet between the scatterings from the two
superconductor-ferromagnet (SF) interfaces~\cite{buzdin_proximity_2005}; the
different orbital modes acquire different phases that add up destructively after
summation. A detailed analysis demonstrates that the proximity effect in a
superconductor-ferromagnet-superconductor (SFS) junction is suppressed
algebraically in the ballistic regime, and exponentially in the diffusive
one~\cite{bergeret_odd_2005}.

Quite recently, it was proposed that a SFS junction with an inhomogeneous
magnetization in the F layer can generate a triplet pairing and can support a
long-range Josephson current~\cite{bergeret_odd_2005}. Subsequent theoretical
and experimental research showed that in order to have dominant triplet
pairing a SFS junction with \textit{two} spin-active regions is
required~\cite{houzet_long_2007,braude_fully_2007,*eschrig_triplet_2008,*asano_josephson_2007,
khaire_observation_2010,volkov_odd_2010,*trifunovic_long-range_2010}.

Motivation for this work is our previous numerical
calculation~\cite{trifunovic_josephson_2011}, where clean and
moderately disordered SFS junctions were considered (with a one spin-active
interface), and the dominant second harmonic was obtained.

In this work, we consider a long SFS junction in the ballistic regime. Assuming
the presence of a spin-active region on only \textit{one} SF interface, we
show that only the phase coherent transport of an \textit{even} number of
Cooper pairs is \textit{not suppressed} by the exchange field. In particular,
the dominant contribution to the Josephson current stems from the transport of
\textit{two} Cooper pairs. As a consequence, the \textit{two} times smaller flux
quantum is obtained, leading to more sensitive quantum interferometers
(SQUIDs)~\cite{radovi_coexistence_2001} and the half-integer Shapiro steps that
can be experimentally observed~\cite{sellier_half-integer_2004}. Another
interesting property is the coexistence of integer and half-integer fluxoid
configurations in SQUIDs, corresponding to the minima of the triple-well
potential energy~\cite{radovi_coexistence_2001}; this can be potentially useful
for experimental study of the quantum superposition of macroscopically distinct
states~\cite{friedman_quantum_2000,*van_der_wal_quantum_2000}. Also, junctions
with a nonsinusoidal current-phase relation are shown to be promising for
realization of ``silent'' phase qubits~\cite{yamashita_superconducting_2005}.
Last but not least, this result enables robust realization of so-called
$\varphi$-junctions~\cite{buzdin_periodic_2003}.

It should be stressed that in contrast to the case of the $0$-$\pi$ transition
the discovered effect is very robust: it is insensitive to a weak
disorder~\cite{trifunovic_josephson_2011}, temperature changes, and the
interface roughness. Nevertheless, relatively transparent interfaces are
required in order to observe the effect.

\begin{figure}
  \includegraphics[width=8cm]{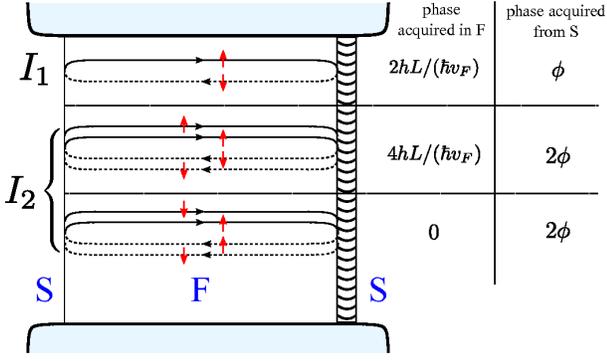}
  \caption{(color online) The first two harmonics in the Josephson current-phase
  relation. The first one ($I_1$) consists of two normal Andreev reflections
  from SF interfaces. The second one ($I_2$) has two contributions: the one with
  four normal Andreev reflections (short-range), the other with two normal and
  two anomalous Andreev reflections (long-range, total phase acquired in the F
  layer is zero). The solid (dotted) lines represent electron (hole) excitations
  in F layer. The red (vertical) arrows represent spin projections, while the
  black (horizontal) arrows denote the excitation velocity direction. The right
  SF boundary (hatched) is spin-active.}
  \label{I_geom}
\end{figure}

Before we proceed with a quantitative analysis of the aforesaid effect, let us
first give a simple and intuitive description. Note that when a SF interface is
spin-active, there are two possibilities for Andreev reflection: the
\textit{normal} Andreev reflection (the spin projections of an electron and the
reflected hole are \textit{opposite}) and the \textit{anomalous} Andreev
reflection (an electron and the reflected hole have the same spin
projections)~\cite{beri_quantum_2009}. We consider separately the phase coherent
transport of one ($I_1$) and two ($I_2$) Cooper pairs. The transport of a single
Cooper pair is suppressed by the exchange field because the electron and the
Andreev reflected hole have opposite spin projections (see Fig. \ref{I_geom}).
On the other hand, the transport of two Cooper pairs has a long-range
contribution stemming from two normal and two anomalous Andreev reflections.

We consider a simple model of a ballistic SFS junction consisting of two
conventional ($s$-wave) superconductors, a uniform single-domain ferromagnet
and only one spin-active region. Andreev reflection requires relatively transparent
SF interfaces, thus for simplicity we assume them to be fully transparent. The
spin-active region consists of a ferromagnetic spacer layer with the
magnetization noncollinear to that of the F layer.

The Josephson current is calculated using the scattering
approach~\cite{brouwer_anomalous_1997}. The knowledge of scattering matrices
(S-matrices) of both SF interfaces is sufficient to obtain the Josephson current
in the ballistic regime. Each of these matrices relates the amplitudes of the
excitations propagating towards the corresponding SF interface to the excitation
propagation away from it. There is no orbital channel mixing in the ballistic
regime, but only mixing of different spin channels (due to spin-active region)
and the particle-hole mixing (due to superconductors). Ergo, the dimension of
the S-matrix is $4\times4$---we write all matrices in the Kronecker product of
particle-hole and spin spaces. The Josephson current is given
by~\cite{brouwer_anomalous_1997}
\begin{equation}
  I=-\frac{2ek_BT}{\hbar}\frac{d}{d\phi}\sum_{n=0}^{\infty}\ln\det[1-{\cal
  R}(i\omega_n){\cal R'}(i\omega_n)],
  \label{Jos_curr}
\end{equation}
where ${\cal R'}$ and ${\cal R}$ are S-matrices of the two SF interfaces, while
the phases acquired upon propagation through the F region are included in one of
these matrices. $\omega_n=(2n+1)\pi k_BT$ are the Matsubara frequencies and
$\phi$ is the phase difference between the two superconductors.

In order to calculate the S-matrix, one has to solve the Bogoliubov-de Gennes
equation for each SF interface. Here we take a simpler approach and express the
S-matrix (${\cal R}$) in terms of a S-matrix of a SN
interface~\cite{beri_quantum_2009,chtchelkatchev_0_2001}, where N stands for
normal-nonferromagnetic layer. By so doing, we neglect the difference in the
number of spin-up or -down modes in the ferromagnet. This approximation is also
assumed in the quasiclassical approach and is justified for a weak exchange
field in the
ferromagnet~\cite{cayssol_exchange-induced_2004,*radovi_josephson_2003}.

The S-matrix of a transparent SN interface reads
\begin{equation}
{\cal R_A}(\varepsilon,\phi)=\alpha(\varepsilon) 
\begin{pmatrix}
0 & i\sigma_2 e^{i \phi}\\
-i\sigma_2 e^{-i \phi} & 0
\end{pmatrix}
,\label{RA}
\end{equation}
where $\alpha(\varepsilon)=\exp{-i\arccos(\varepsilon/\Delta)}$ and $\sigma_2$
is the second Pauli matrix. For the ferromagnetic spacer layer (spin-active
region), the S-matrix is
\begin{equation}
S_F=
\begin{pmatrix}
  0 & U \\
  U & 0
\end{pmatrix}
,\quad U=\exp{i(\eta+\rho\bm m\cdot\bm\sigma)/2},
\label{SF}
\end{equation}
where $\bm m=(\sin\theta\cos\psi,\sin\theta\sin\psi,\cos\theta)$ is the
magnetization orientation in the ferromagnetic spacer layer. The difference of
the phase shifts of spin-up and spin-down electrons upon propagating through the
ferromagnetic spacer layer is denoted by $\rho=\nu_\uparrow-\nu_\downarrow$,
while $\eta=\nu_\uparrow+\nu_\downarrow$. Here the spin-up (spin-down) is
defined with respect to the magnetization axis in the F layer. These phases
depend on the orbital channel index $\mu$, but for the sake of notational
simplicity we have suppressed the index. We also assume that the ferromagnetic
spacer layer thickness is much smaller than the superconducting coherence length
($L'\ll\xi_S$), so that the energy dependence of $\rho$ and $\eta$ can be
ignored. Combining these scattering matrices we obtain the scattering matrix of
the SF interface with the spin-active region
\begin{equation}
  {\cal R}(\varepsilon)=\alpha(\varepsilon)
  \begin{pmatrix}
    0 & -\hat r_{he}^*\exp{i\phi}\\
    \hat r_{he}\exp{-i\phi} & 0
  \end{pmatrix},
  \label{Rmatrix}
\end{equation}
with
\begin{equation}
  \hat r_{he}=
  \begin{pmatrix}
    -i\exp{i\psi}\sin\theta\sin\rho &
    -\cos\rho+i\cos\theta\sin\rho\\
    \cos\rho+i\cos\theta\sin\rho &
    i\exp{-i\psi}\sin\theta\sin\rho
  \end{pmatrix}.
  \label{rhe}
\end{equation}

For the SF interface without the spin-active region, we also include the phases
acquired in the F layer and obtain
\begin{equation}
  {\cal R'}(\varepsilon)=T(\varepsilon){\cal
R_A}(\varepsilon,0)T(\varepsilon),
  \label{Rpmatrix}
\end{equation}
with $T=\exp{i{\rm
diag}[k_{\mu,\uparrow}(\varepsilon),k_{\mu,\downarrow}(\varepsilon),
-k_{\mu,\uparrow}(-\varepsilon),-k_{\mu,\downarrow}(-\varepsilon)]L}$;
$k_{\mu,\uparrow\downarrow}$ is the longitudinal component of wavevector in the
orbital mode $\mu$ (spin-up/down), and $L$ is the F layer thickness.

In order to perform the integration over orbital modes, we put
$\rho_\mu=Z'/\cos\Theta$, where $Z'=2h'L'/(\hbar v_F)$; $h'$ is the exchange
energy in the spacer layer which is assumed to be small ($h'\ll E_F$), and
$\Theta$ is the angle between the excitation velocity and the junction axis.
Also, in the Andreev approximation
$k_{\mu,\uparrow\downarrow}(\varepsilon)=(k_F\pm h/\hbar v_F+\varepsilon/\hbar
v_F)/\cos\Theta$, where $h$ is the exchange energy in the F layer ($h\ll E_F$).
Setting $L'=0$ (no spin-active layer) recovers the result previously obtained
from the quasiclassical approach~\cite{radovi_coexistence_2001}, where the
following expression relates the scattering to the quasiclassical approach
(see also Ref. \onlinecite{kupferschmidt_andreev_2011})
\begin{equation}
  \sum_{\sigma}\frac{g_\sigma(\Theta)-g_\sigma(-\Theta)}{2}=
  \frac{e^2R_N}{i\pi\hbar}\frac{d}{d\phi}\ln\det(1-{\cal R}{\cal R'}),
  \label{quasi_scatt}
\end{equation}
where $R_N$ is the normal resistance and $g_\sigma$ is the normal Green function
in the ferromagnet~\cite{radovi_coexistence_2001}.

In the general case---with one spin-active region---upon inserting Eqs.
(\ref{Rmatrix}), (\ref{Rpmatrix}) into Eq. (\ref{Jos_curr}) the formula for the
Josephson current is obtained. We introduce a new variable
$u=\arcsinh(\omega/\Delta)$, and integrate over orbital modes 
\begin{align}
  \label{Jos_inter}
  I=&\frac{2\pi T}{eR_N}\sum_{\sigma,\omega_n>0} \int_0^{\pi/2}
  d\Theta\sin\Theta\cos\Theta \\
  &\times\Im \tanh \left[u + \Delta \sinh u
  \frac{L+L'}{\hbar v_F \cos \Theta} + i\sigma\frac{\chi}{2} + i
  \frac{\varphi}{2} \right],
  \nonumber
\end{align}
with $\chi=\arccos(\Re[r_{he}^{\downarrow\uparrow}\exp{-iZ/\cos\Theta}])$ and
$Z=2hL/(\hbar v_F)$; $r_{he}^{\downarrow\uparrow}$ denotes the element $(2,1)$
of the matrix in Eq. (\ref{rhe}).

We now consider a long SFS junction, and show that only even harmonics in the
CPR are long-range. In this case ($L\gg\xi_S$), the first term in the argument
of the hyperbolic tangent ($u$) can be neglected. At zero temperature, the
summation over $\omega_n$ can be replaced by an integration; the expression for
the Josephson current reads
\begin{equation}
  I=\frac{4\hbar v_F}{eR_NL}\sum_{k=1}^{\infty}(-1)^kI_k\sin(k\phi).
  \label{Ilong}
\end{equation}
In the last equation, the spectral weights $I_k$ quantify the contribution to
the current coming from the phase coherent transport of $k$ Cooper pairs across
the barrier
\begin{equation}
  I_k=-\int_1^\infty\frac{T_k(\Re[r_{he}^{\downarrow\uparrow}\exp{-iZx}])/k}{x^4}dx,
  \label{Ik}
\end{equation}
where $T_k$ is the Chebyshev polynomial of the first kind and $x=1/\cos\Theta$.
We have used the identity $\Im\tanh
z=2\sum_{k=1}^\infty(-1)^k\Im\exp{-2kz}$ for obtaining Eqs.
(\ref{Ilong}-\ref{Ik})~\cite{svidzinskii_spatially_1982}.

In order to avoid cumbersome expressions, we concentrate on the case with
mutually orthogonal magnetizations in the F and the ferromagnetic spacer layer
($\theta=\pi/2$). Later we will show that all our conclusions are valid for
arbitrary (but not too small) $\theta$. Now,
$r_{he}^{\downarrow\uparrow}=\cos(Z'x)$ and the expression for the spectral
weights reads
\begin{align}
  \label{Ijongpi2}
  I_k=&\int_1^\infty\frac{dx}{x^4}\sum_{n=0}^{\lfloor
  k/2\rfloor}\sum_{m=0}^{n}\frac{(-1)^{n+m+1}}{k}\binom{k}{2n}\binom{n}{m}\\
  &\times[\cos(Zx)\cos(Z'x)]^{k-2(n-m)},\nonumber
\end{align}
where $\lfloor x\rfloor$ denote the largest integer not greater than $x$.
\begin{figure}
  \includegraphics[width=8cm]{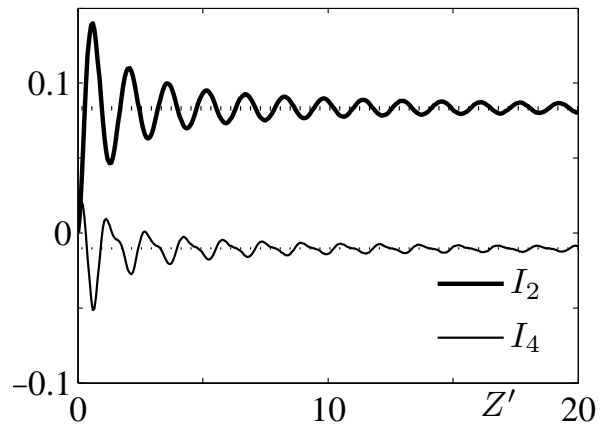}
  \caption{The long-range component of the second (thick solid curve) and the
  fourth (thin solid curve) harmonic of the Josephson current-phase relation.
  For $Z'\gg1$, both the curves converge to constant values (dotted lines),
  $\bar I_2$ and $\bar I_4$.}
  \label{I_12}
\end{figure}

We perform the integration in Eq. (\ref{Ijongpi2}) by expanding the integrand as
a sum of cosines. When $k$ is \textit{odd}, every term in that sum
\textit{depends} on $Z$; for a long SFS junction and for a reasonably strong
exchange energy ($h>\Delta$) we have $Z\gg1$ and
$\int_1^\infty\cos(Zx)/x^4dx=-\sin Z/Z+O(1/Z^2)$. Hence, odd harmonics are
suppressed by the factor $1/Z$. On the other hand, for \textit{even} $k$ we find
terms that are \textit{independent} of $Z$ (the exchange energy in the F
layer). Thus we write $I_{2k}=I_{2k}^{LR}+O(1/Z)$. The long-range component of
even harmonics ($I_{2k}^{LR}$) is given by
\begin{align}
    \label{Ijongpi2LR}
    I_{2k}^{LR}=&\int_1^\infty\frac{dx}{x^4}\sum_{n=0}^{k}\sum_{m=0}^{n}
    \frac{(-1)^{n+m+1}}{2^{2(k-n+m)}}\frac{1}{2k}\\
    &\times\binom{2(k-n+m)}{k-n+m}\binom{2k}{2n}\binom{n}{m}
    \cos(Z'x)^{2(k-n+m)}.\nonumber
\end{align}
This result shows that \textit{even} harmonics dominate in a long junction. It
should be noted that $even$ harmonics can dominate for any junction length if
the parameters of the spin-active region are chosen in such a way that normal
Andreev reflection vanishes on one SF interface (i.e.
$r_{he}^{\downarrow\uparrow}=0$). This can be directly seen from Eqs.
(\ref{Ilong}-\ref{Ik}).

\begin{figure}
  \includegraphics[width=8cm]{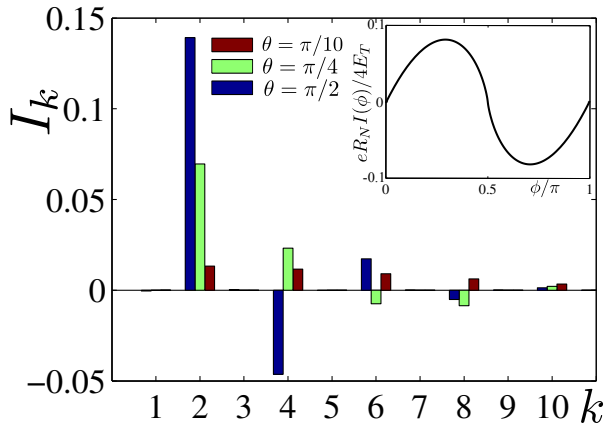}
  \caption{(color online) The spectral weights of the current-phase relation
  for $Z'=0.5$ and $Z=1000$, and for three values of the relative angle between
  magnetizations: $\theta=\pi/10,\;\pi/4,\;\pi/2$. The inset depicts the generic
  CPR given by Eq. (\ref{Igeneric}). The Josephson current is normalized by
  $eR_N/(4E_T)$, where $E_T=\hbar v_F/L$ is the Thouless energy.}
  \label{I_theta}
\end{figure}

Figure \ref{I_12} shows the dependence of the first two long-range harmonics
($I_2$ and $I_4$) on $Z'$. For large values of $Z'$ both curves converge to
the constant values---$\bar I_2$ and $\bar I_4$, respectively. Hence, we see
that for a certain range of the ferromagnetic spacer layer parameters [$1\ll
L'h'/(\hbar v_F)\ll h'/\Delta$] the Josephson current reads
\begin{equation}
  I=\frac{4\hbar v_F}{eR_NL}\sum_{k=1}^{\infty}\bar
  I_{2k}^{LR}\sin(2k\phi)+O\left(  \frac{\hbar v_F}{hL},\frac{\hbar
  v_F}{h'L'}\right),
    \label{Igeneric}
\end{equation}
where $\bar I_{2k}^{LR}$ are independent of the junction parameters
\begin{align}
    \label{Ibar}
    \bar I_{2k}^{LR}=&\sum_{n=0}^{k}\sum_{m=0}^{n}
    \frac{(-1)^{n+m+1}}{2^{4(k-n+m)}}\frac{1}{6k}\\
    &\times\binom{2(k-n+m)}{k-n+m}^2\binom{2k}{2n}\binom{n}{m}.\nonumber
\end{align}
Equations (\ref{Igeneric}-\ref{Ibar}) assert that for a long SFS junction with
one spin-active region, the long-range part of the Josephson current depends
only on the Thouless energy. Consequently, the long-range part of the current is
not suppressed by fluctuations of the ferromagnetic barrier thickness (interface
roughness). The current-phase relation is depicted in the inset of Fig.
\ref{I_theta}.

In the general case, when the value of $Z'$ is arbitrary (but $L'\ll\xi_S$), the
supercurrent dependence on $Z'$ is given by Eq. (\ref{Ijongpi2LR}). Again, even
harmonics are dominant. The free energy of Josephson junction is given by
$F(\phi)\propto\int_0^\phi I(\tilde\phi)d\tilde\phi$; the ground state of the
junction is degenerate: the $0$ and $\pi$ states have the same energy. The
degeneracy is lifted only by algebraically small factors ($\sim1/Z$).

The dependence of the spectral weights on the misorientation angle is depicted
for three values of $\theta$ in Fig. \ref{I_theta}. We observe that
\textit{even} harmonics dominate also for $\theta<\pi/2$. As the relative angle
between magnetizations approaches zero, the amplitudes of even harmonics lessen
and eventually the odd and even harmonics become comparable. Therefore, one can
tune the ratio $I_2/I_1$ by changing the angle $\theta$, while the sign of $I_1$
can be changed by adjusting the thickness of the F
layer~\cite{chtchelkatchev_0_2001}. This enables robust realization of
$\varphi$-junctions~\cite{buzdin_periodic_2003}, contrary to previously
discussed realizations with $0$-$\pi$ transition.

At finite temperatures a new length scale appears---the normal metal coherence
length $\xi_N=\hbar v_F/(2\pi T)$. For high temperatures, $T\gg\hbar v_F/L=E_T$
(i.e. $L\gg\xi_N$), only the first term ($n=0$) in the summation over Matsubara
frequencies contributes to the current given by Eq. (\ref{Jos_inter}). After
performing the integration over orbital modes, assuming $h\gg T$, we obtain
\begin{align}
    I=&-\frac{8\pi T}{eR_N}\left[
    \frac{\sin(Z-Z')}{Z}\cos^2\frac{\theta}{2}+\frac{\sin(Z+Z')}{Z}\sin^2\frac{\theta}{2}
    \right]\nonumber\\
    &\times\frac{\Delta^2}{(\pi
    T+\sqrt{\Delta^2+\pi^2T^2})^2}\exp{-L/\xi_N}\sin\phi\:.
    \label{IhighT}
\end{align}
We conclude that the \textit{first} harmonic always dominate in the
\textit{high}-temperature limit, because the higher harmonics are
suppressed by the factor $\exp{-k\xi_N/L}$ ($k>1$)~\footnote{\textit{Higher
harmonics} originate from the phase coherent transport of \textit{more than one}
Cooper pair across the barrier. Such coherent transport becomes highly
suppressed when the coherence length is much smaller then the barrier thickness
($\xi_N\ll L$).}; in this case the supercurrent has only the short-range
part~\cite{houzet_long_2007}.

In conclusion, we have shown that SFS junctions with \textit{one} and
\textit{two} spin-active SF interfaces are qualitatively different. In the
case of \textit{two} spin-active interfaces, \textit{all} harmonics in the
Josephson current-phase relation are long-range (and the first one is
dominant)~\cite{volkov_odd_2010}, while in the case of \textit{one} spin-active
interface, we find that only $even$ harmonics are long-range (and the second one
is dominant). Some repercussions of the discovered effect are: half-integer
Shapiro steps~\cite{sellier_half-integer_2004}, the coexistance of integer and
half-integer fluxoid SQUID configurations~\cite{radovi_coexistence_2001} and
robust realization of the $\varphi$-junctions~\cite{buzdin_periodic_2003}.

I acknowledge C.~Bruder and C.W.J~Beenakker for reading the manuscript. I am
especially grateful to Z.~Radovi\'c and V.~M.~Stojanovi\'c for many useful
discussions. This work is supported by the Swiss NSF, NCCR Nanoscience (Basel),
and DARPA QuEST.
\bibliography{condmat}
\end{document}